\documentclass[fleqn,11pt]{article}
\usepackage{amscd,amsmath,amssymb,verbatim}
\usepackage{amsthm}
\usepackage[pdftex]{graphicx}
\usepackage{epstopdf}
\usepackage[T1]{fontenc}
\usepackage{ae,aecompl}
\usepackage{subfigure}
\usepackage{array}
\usepackage{mathrsfs}
\usepackage[nohead]{geometry}
\usepackage[singlespacing]{setspace}
\usepackage{rotating}
\usepackage{multirow}
\usepackage{threeparttable}
\usepackage{enumerate}
\usepackage{epstopdf}
\usepackage{subfigure}
\usepackage{placeins}
\usepackage{color, colortbl}
\usepackage{accents}
\usepackage{pdflscape}
\usepackage{natbib}
\usepackage{float}
\numberwithin{equation}{section}

\DeclareMathOperator{\Diag}{Diag}

\newcommand{\bs}{\boldsymbol}
\newcommand{\E}{\mathbb{E}}

\newcommand{\var}{\mathrm{Var}}
\newcommand{\mb}{\mathbb}
\newcommand{\mf}{\mathbf}

\newcommand{\h}{\mathbf{h}}
\newcommand{\Hh}{\mathbf{H}}
\newcommand{\W}{\mathbf{W}}

\newcommand{\U}{\mathbf{U}}
\newcommand{\V}{\mathbf{V}}
\newcommand{\I}{\mathbf{I}}
\newcommand{\Y}{\mathbf{Y}}

\newcommand{\Ss}{\mathbf{S}}
\newcommand{\A}{\mathbf{A}}
\newcommand{\J}{\mathbf{J}}
\newcommand{\K}{\mathbf{K}}
\newcommand{\Pp}{\mathbf{P}}
\newcommand{\0}{\mathbf{0}}

\newcommand{\Lag}{\mathbf{L}}
\newcommand{\B}{\mathbf{B}}

\newcommand{\Z}{\mathbf{Z}}

\newcommand{\la}{\boldsymbol{\lambda}}
\newcommand{\rh}{\boldsymbol{\rho}}

\newcommand{\Om}{\boldsymbol{\Omega}}
\newcommand{\Si}{\boldsymbol{\Sigma}}

\usepackage{hyperref}

\makeatletter
\newcommand{\distas}[1]{\mathbin{\overset{#1}{\kern\z@\sim}}}%
\newsavebox{\mybox}\newsavebox{\mysim}
\newcommand{\distras}[1]{%
  \savebox{\mybox}{\hbox{\kern3pt$\scriptstyle#1$\kern3pt}}%
  \savebox{\mysim}{\hbox{$\sim$}}%
  \mathbin{\overset{#1}{\kern\z@\resizebox{\wd\mybox}{\ht\mysim}{$\sim$}}}%
}
\makeatother

\newtheorem{algorithm}{Algorithm}

\newtheorem{rmk}{Remark}

\renewcommand{\hat}[1]{\widehat{\text{$#1$}}}
\makeatletter
\newsavebox\myboxA
\newsavebox\myboxB
\newlength\mylenA

\renewcommand*\bar[2][0.85]{%
    \sbox{\myboxA}{$\m@th#2$}%
    \setbox\myboxB\null
    \ht\myboxB=\ht\myboxA%
    \dp\myboxB=\dp\myboxA%
    \wd\myboxB=#1\wd\myboxA
    \sbox\myboxB{$\m@th\overline{\copy\myboxB}$}
    \setlength\mylenA{\the\wd\myboxA}
    \addtolength\mylenA{-\the\wd\myboxB}%
    \ifdim\wd\myboxB<\wd\myboxA%
       \rlap{\hskip 0.5\mylenA\usebox\myboxB}{\usebox\myboxA}%
    \else
        \hskip -0.5\mylenA\rlap{\usebox\myboxA}{\hskip 0.5\mylenA\usebox\myboxB}%
    \fi}
\makeatother
\hoffset 
\voffset
\author{ Philipp Otto\thanks{Leibniz University, Hannover, Germany, email: philipp.otto@ikg.uni-hannover.de.} \and Osman Do\u{g}an\thanks{Economics Department, Istanbul Technical University, Istanbul, Turkey, email: odogan10@gmail.com.}\and S\"uleyman Ta\c{s}p{\i}nar\thanks{Department of Economics, Queens College, The City University of New York, United States, email: staspinar@qc.cuny.edu.}}\medskip

\date{\today}
\title{A Dynamic Spatiotemporal Stochastic Volatility Model with an Application to Environmental Risks}

\begin{document}
\maketitle
\sloppy

\singlespacing

\begin{abstract}
\noindent 
This article introduces a dynamic spatiotemporal stochastic volatility (SV) model with explicit terms for the spatial, temporal, and spatiotemporal spillover effects. Moreover, the model includes time-invariant site-specific constant log-volatility terms. Thus, this formulation allows to distinguish between spatial and temporal interactions, while each location may have a different volatility level. We study the statistical properties of an outcome variable under this process and show that it introduces spatial dependence in the outcome variable. Further, we present a Bayesian estimation procedure based on the Markov Chain Monte Carlo (MCMC) approach using a suitable data transformation. After providing simulation evidence on the proposed Bayesian estimator's performance, we apply the model in a highly relevant field, namely environmental risk modeling. Even though there are only a few empirical studies on environmental risks, previous literature undoubtedly demonstrated the importance of climate variation studies. For example, for local air quality in Northern Italy in 2021, we show pronounced spatial and temporal spillovers and larger uncertainties/risks during the winter season compared to the summer season.
\end{abstract}
\noindent
JEL-Classification: C13, C21, C31.

\onehalfspacing

\section{Introduction}\label{intro}

When analyzing geo-referenced data, statistical models have to account for instantaneous spatial correlations due to the geographical proximity between the observations. This is commonly known as Tobler's first law of geography ``everything is related to everything else, but near things are more related than distant things'' \citep{Tobler70}. This observation was already noted by Ronald A. Fisher in 1935 as follows, ``the widely verified fact that patches in close proximity are commonly more alike, as judged by the yield of crops, than those which are further apart'' \citep{fisher1935design}. Even though the similarity is typically considered to be in the (conditional) mean level at each location, there might also be spatial correlations in the (conditional) variance or variation of the random process. In particular, for small-scale spatial units, the variance of the process is increased \citep[known as Arbia's law of geography,][]{arbia1996effects}. In addition to the instantaneous spatial correlations, we also have to account for the natural temporal correlations, which usually occur if we repeatedly observe a random process over time. The closer two observations are to each other in time, the more strongly they can correlate in general. In this paper, we introduce a new model for spatial, temporal, and spatiotemporal correlations in the log-volatilities allowing for additional random errors in the mean and volatility equation. Furthermore, we apply this model to environmental data and show for the first time how it can be used to analyze environmental risk factors such as air pollution. 

There are generally two ways to account for spatial/cross-sectional correlations in spatial statistics. Firstly, it can be modeled in the covariance matrix of the process, where each entry is supposed to follow a certain (non-)parametric covariance function depending on the distance between their locations. This idea is typically known as the geostatistical approach \citep{cressie2015statistics, zimmerman2014geostatistics}. The selection of a suitable parametric covariance function with certain properties such as stationarity, separability and full symmetry is one of the main modeling issues of this approach \citep{Porcu16,huang2011class}. See \citet{Gneiting:2007} for a review on the spatiotemporal covariance functions suggested in the literature. Secondly, the observations on an outcome variable can be explicitly correlated with the adjacent observations, where the adjacency is defined fairly generally by a spatial weights matrix. This second approach is referred to as spatial autoregression in spatial econometrics, where the spatial lags of variables are used to model spatial correlations. See \citet{Lesage:2009, Anselin:1988, Elhorst:2014, Lee:2004, KP:2010} on the specification and estimation issues in spatial econometrics. Both approaches can be equivalent under certain conditions \citep[see e.g.][for simultaneous and conditionally autoregressive models]{ver2018relationship}. 

In this paper, we consider the second approach. Our model consists of an outcome and log-volatility equation with separate independent error terms, whereby the log-volatility process introduces spatial dependence in the outcome variable. Specifically, the log-volatility equation allows for spatial, temporal, and spatiotemporal correlations, as well as time-invariant site-specific effects (unobserved heterogeneity). Also, assuming that the error terms in both equations have normal distributions, it is possible to show that the outcome variable has a leptokurtic symmetric distribution under our suggested model. To introduce a Bayesian estimation approach, we use a transformation approach such that the outcome equation becomes linear in the log-volatility terms. We use a Gaussian mixture distribution to approximate the distribution of the transformed error terms in the outcome equation. This approximation turns our model into a linear state-space model, where the log-volatility equation becomes the state equation. Following recent developments in  the precision-based algorithms \citep{Chan:2009, Chan:2017}, we suggest a Gibbs sampler that consists of five steps for the estimation. We provide simulation evidence showing that the suggested sampler can perform satisfactorily. 

Theoretically, our paper is related to the spatial econometric literature that addresses the presence of cross-sectional correlations in higher moments of spatial data. This strand of the literature considers the spatial extensions of generalized autoregressive conditional heteroskedasticity (GARCH) and stochastic volatility (SV) models to account for volatility clustering patterns observed over space \citep{Otto:2018, Hol:2020, Takaki:2021, Taspinar:2021, Robinson:2009,Yan:2007}. Our model can be considered as the  longitudinal data (panel data) extension of the cross-sectional spatial SV models suggested by  \citet{Yan:2007, Taspinar:2021, Robinson:2009}. While these studies allow for the presence of spatial dependence in the log-volatility equations, they do not include temporal, spatiotemporal and unobserved heterogeneity terms in the log-volatility equations. Our suggested process is also related to the separable and non-separable space-time filters considered in the spatial panel data models for modeling spatiotemporal interactions. For example, \citet{Lesage:2012, Lesage:2011} consider a separable space-time filter for the outcome variable, while \citet{Yu:2015} and \citet{Wang:2018} consider non-separable space-time filters for the disturbance terms of spatial panel data models. In contrast to these studies, we consider a general space-time filter that also allows for unobserved heterogeneity, i.e., the site-specific effects, for the log-volatility of an outcome variable.   

In an empirical application, we use our suggested model to assess environmental risk stemming from the variations in the log-volatility of air quality predictions. Spatial and spatiotemporal interactions in the local climate and environmental risks were addressed in comparably few studies in previous empirical research, even though it has been shown that an increased variation in environmental processes can be harmful \citep{tewksbury2008putting, paaijmans2013temperature, vasseur2014increased, Iaco12}. Most previous studies focused on correlations in the (conditional) mean levels of ecological processes \citep[e.g.]{ver2018spatial,wilby2009review}. In our case, we model the log-volatility of fine dust concentrations of particles having a diameter less than $10\mu m$ in Lombardy, Northern Italy. Following the literature on the ecological processes \citep[e.g.]{ver2018spatial,wilby2009review}, we first model the variation in the conditional mean of our outcome variable through a conventional spatial panel data model that allows for the unobserved site and time heterogeneity. Our results from this initial model indicate that there is strong and moderate spatial correlations in the outcome variable in the model with only site fixed effects and the model with both site and time fixed effects, respectively. In the next step, we use the errors from this initial model as an outcome variable in our suggested specification and aim to model the variations in its log-volatility terms. The estimation results from our suggested Bayesian approach indicate that the spatial and temporal effects in the log-volatilities are moderate while the spatiotemporal effects appear to be of minor importance. We were also able to detect a noticeable variation in air pollution risk across the year and identify measurement stations that are associated with higher risks. These are mostly located in valleys in the Alpine regions.


The rest of this paper proceeds in the following way. In Section~\ref{spec}, we introduce our suggested model specification, including conditions ensuring the stability of the model and prior specifications. In Section~\ref{prop}, we investigate the statistical properties of the suggested model. In Section~\ref{post}, we provide the details on the posterior analysis of our model, and state an algorithm for the estimation. Section~\ref{sim} provides a simulation study on the performance of the suggested Gibbs sampler. Then, in Section~\ref{emp}, such stochastic volatility model is applied to environmental risks for the first time. More precisely, our focus in on local air quality modeling. Finally, in Section~\ref{conc}, we provide our concluding remarks.

\section{Model Specification}\label{spec}

Suppose that we observe the spatiotemporal process across a constant set of $n$ locations in a geographical domain at $T$ equidistant time points. These locations can be measurement stations (i.e., marked point data), atmospheric and remotely-sensed data, or sets of municipalities, counties, states (i.e., areal data). Moreover, these locations does not have to seen in a strict geographical sense, but can also be vertices in a network. Let $\Y_{t}=(y_{1t}, y_{2t}, \hdots,y_{nt})^{'}$ be the $n\times 1$ vector of the outcome variable at time $t$ for $t=1,2,\hdots,T$. We assume the following data generating process (DGP) for $\Y_t$:
\begin{align}\label{2.1}
&\Y_{t}=\Hh^{1/2}_t\V_{t}, 
\end{align}
for $t=1,2,\hdots T$, where $\Hh^{1/2}_t=\Diag(e^{\frac{1}{2}h_{1t}},\hdots,e^{\frac{1}{2}h_{nt}})$ is the $n\times n$ diagonal matrix containing stochastic volatility terms $h_{it}$'s, which are specified subsequently, and $\V_{t}=(v_{1t},\hdots,v_{nt})^{'}$ is the $n\times 1$ vector of disturbance terms. We assume that $v_{it}$'s  are i.i.d standard normal random variables.  Let $\h_t=(h_{1t},\hdots,h_{nt})^{'}$ be the $n\times1$ vector of stochastic volatility at time $t$. We assume the following process for $\h_t$:
\begin{align}\label{2.2}
&\h_t-\bs{\mu}=\rho_1\W(\h_t-\bs{\mu})+\rho_2(\h_{t-1}-\bs{\mu})+
\rho_{3}\W(\h_{t-1}-\bs{\mu})+\U_t,
\end{align}
for $t=1,2,\hdots T$, where $\bs{\mu}=(\mu_1,\hdots,\mu_n)^{'}$ is the $n\times1$ vector of constant means, i.e., the time-invariant site-specific effects,  and $\U_{t}=(u_{1t},\hdots,u_{nt})^{'}$ is the $n\times1$ vector of i.i.d. disturbance terms such that $u_{it}\sim N(0, \sigma^2)$ for all $i$ and $t$. In \eqref{2.2}, $\W$ is the $n\times n$ spatial weights matrix that has zero diagonal elements. This matrix specifies how volatility terms are related over space. In a network setting, this matrix is equivalently specified as an adjacency matrix. The scalar parameter $\rho_1$ captures contemporaneous spatial correlation, $\rho_2$ measures the temporal effect, i.e., the time dynamic effect, and $\rho_3$ represents the spatiotemporal effect, i.e., the spatial diffusion effect. Let $\Lag$ be the matrix of time-lag operator such that $\Lag\h_t=\h_{t-1}$. Then, \eqref{2.2} can be written as 
\begin{align}\label{3.3}
\left((\I_n-\rho_1\W)-(\rho_2\Lag+\rho_3\W\Lag)\right)(\h_t-\bs{\mu})=\U_t,
\end{align}
where $\left((\I_n-\rho_1\W)-(\rho_2\Lag+\rho_3\W\Lag)\right)$ is called the general space-time filter \citep{Lesage:2011, Lesage:2012,Yu:2015}. Under the assumption that $\rho_3=-\rho_1\rho_2$, this general filter is separable and decomposes into a product of the space filter $(\I_n-\rho_1\W)$ and the time filter $(\I_n-\rho_2\Lag)$. In our analysis, we do not impose this restrictive assumption. Let $\Ss(\rho_1)=(\I_n-\rho_1\W)$. Then, under the assumption that $\Ss(\rho_1)$ is invertible, the reduced form of volatility equation is
\begin{align}
&\h_t-\bs{\mu}=\Ss^{-1}(\rho_1)\A(\rho_2,\rho_3)(\h_{t-1}-\bs{\mu})+
\Ss^{-1}(\rho_1)\U_t,\label{3.5}
\end{align}
where $\A(\rho_2,\rho_3)=(\rho_2\I_n+\rho_3\W)$. When the cross-sectional dimension is fixed, the process for the log-volatility is stable if all eigenvalues of $\Ss^{-1}(\rho_1)\A(\rho_2,\rho_3)$ lie inside the unit ball \citep[Proposition 10.1]{Hamilton:1994}. Let $\vartheta_i(\W)$ be the $i$th eigenvalue of $\W$ for $i=1,\hdots,n$. We assume that the parameter space of $\rho_1$, $\rho_2$ and $\rho_3$ are chosen such that the following conditions hold:\footnote{In the spatial econometric literature, there are alternative ways to specify the parameter spaces for spatial autoregressive parameters. Among others, see \citet{Anselin:1988}, \citet{Lesage:2009}, \citet{Lee:2004}, \citet{KP:2010} and \citet{Elhorst:2014}.}
\begin{align}\label{3.6}
(i) \max_{1\leq i\leq n}\left|\vartheta_i(\rho_1\W)\right|<1,\quad \text{and} \; (ii)\,\max_{1\leq i\leq n}\left|\vartheta_i\left(\Ss^{-1}(\rho_1)\A(\rho_2,\rho_3)\right)\right|<1.
\end{align}
The first condition is the sufficient condition for the invertibility of $\Ss(\rho_1)$ \citep[Lemma 1]{KP:2010}. By the spectral radius theorem \citep[Theorem 5.6.9]{Horn:2012}, we can use any matrix norm to define relatively restrictive conditions that ensure the conditions in \eqref{3.6}. Let $\Vert\cdot\Vert$ be any matrix norm. Then, the sufficient conditions for  \eqref{3.6} are (i) $\Vert\rho_1\W\Vert<1$ and (ii) $\Vert\Ss^{-1}(\rho_1)\A(\rho_2,\rho_3)\Vert<1$. Note that 
\begin{align}\label{3.7}
\left\Vert \Ss^{-1}(\rho_1)\A(\rho_2,\rho_3)\right\Vert&\leq\left\Vert \Ss^{-1}(\rho_1)\right\Vert\times\left\Vert \rho_2\I_n+\rho_3\W\right\Vert\nonumber\\
&=\Vert\I_n+\rho_1\W+\rho^2_1\W^2+\rho^3_1\W^3+\hdots\Vert\times\Vert \rho_2\I_n+\rho_3\W\Vert\nonumber\\
&\leq\left(\Vert \I_n\Vert+\Vert\rho_1\W\Vert+\Vert\rho_1\W\Vert^2+\hdots\right)
\times\left(|\rho_2|+|\rho_3|\cdot\Vert\W\Vert\right)\nonumber\\
&=\frac{1}{1-\Vert\rho_1\W\Vert}\times\left(|\rho_2|+|\rho_3|\cdot\Vert\W\Vert\right),
\end{align}
where last equality follows since we assume that $\Vert\rho_1\W\Vert<1$. If we choose the matrix row sum norm $\Vert\cdot\Vert_{\infty}$ and assume that $\W$ is row normalized, then \eqref{3.7} reduces to $(|\rho_2|+|\rho_3|)/(1-|\rho_1|)$. Thus, a further restrictive sufficient condition for the stability of \eqref{3.5} is $|\rho_1|+|\rho_2|+|\rho_3|<1$. We will impose these restrictions during the sampling steps for $\rho_1$, $\rho_2$ and $\rho_3$ in our suggested Gibbs sampler. 

Finally, to complete the model in \eqref{2.1}, we assume the following prior distributions for the posterior analysis:
\begin{align}\label{3.8}
&\rho_1\sim\text{Uniform}(-1,1),\quad\rho_2\sim\text{Uniform}(-1,1),\,\rho_3\sim\text{Uniform}(-1,1),\\ 
&\bs{\mu}|\mf{b}_{\mu},\B_{\mu}\sim N(\mf{b}_{\mu},\B_{\mu}),\,\sigma^2|a,b\sim\text{IG}(a,b),\nonumber
\end{align}
where $\text{Uniform}(c_1,\,c_2)$ denotes the uniform distribution over the interval $(c_1,\,c_2)$ and $\text{IG}(a,b)$ denotes the inverse gamma distribution with the shape parameter $a$ and the scale parameter $b$. The priors for $\rho_1$, $\rho_2$ and $\rho_3$ are subject to the stability conditions sated in \eqref{3.6}.

\section{Statistical Properties}\label{prop}

The outcome equation of our model can be written as 
\begin{align}\label{4.1}
y_{it}=e^{\frac{1}{2}h_{it}}v_{it} \qquad \text{for all $i = 1, \ldots, n$ and $t = 1, \ldots, T$.}
\end{align}
Thus, the conditional variance of $y_{it}$ given $h_{it}$ is $\text{Var}(y_{it}|h_{it})=e^{h_{it}}$, indicating that the conditional variance is both time and space varying. Following the time series literature, we refer to $h_{it}$ as the log-volatility since $h_{it}=\log\left(\text{Var}(y_{it}|h_{it})\right)$. In order to determine the unconditional moments of $y_{it}$, we need to determine the distribution of $\h=(\h^{'}_1,\hdots,\h^{'}_T)^{'}$. Let $\rh=(\rho_1,\rho_2,\rho_3)^{'}$, and define the $nT\times nT$ matrix $\J(\rh)$ as
\begin{align}\label{4.9}
\J(\rh)=
\begin{pmatrix}
\Ss(\rho_1)&\0&\hdots&\0&\0\\
-\A(\rho_2,\rho_3)&\Ss(\rho_1)&\hdots&\0&\0\\
\vdots&\ddots&\ddots&\vdots&\vdots\\
\0&\0&\hdots&-\A(\rho_2,\rho_3)&\Ss(\rho_1)
\end{pmatrix}.
\end{align}
Then, we can express the log-volatility equation as  
\begin{align}\label{4.10}
\J(\rh)(\h-\bs{l}_T\otimes\bs{\mu})=
\begin{pmatrix}
\Ss(\rho_1)(\h_1-\bs{\mu})\\
\U_2\\
\vdots\\
\U_T\\
\end{pmatrix}.
\end{align}
We assume
that the spatial dynamic process for $\mathbf{h}_t$ has been operating for a long time so that we can express $\Ss(\rho_1)(\h_1-\bs{\mu})$ in the following way:
\begin{align}\label{recursive}
\Ss(\rho_1)(\h_1-\bs{\mu})=\sum_{j=0}^{\infty}\left(\A(\rho_2,\rho_3)\Ss^{-1}(\rho_1)\right)^{j}\U_{1-j},
\end{align}
which implies that 
\begin{align*}
\var(\Ss(\rho_1)(\h_1-\bs{\mu}))&=\sigma^2\sum_{j=0}^{\infty}\left(\A(\rho_2,\rho_3)\Ss^{-1}(\rho_1)\right)^{j}\left(\A(\rho_2,\rho_3)\Ss^{-1}(\rho_1)\right)^{'j}=\sigma^2\K(\rh),
\end{align*}
where $\K(\rh)=\sum_{j=0}^{\infty}\left(\A(\rho_2,\rho_3)\Ss^{-1}(\rho_1)\right)^{j}\left(\A(\rho_2,\rho_3)\Ss^{-1}(\rho_1)\right)^{'j}$. Then, from \eqref{4.10}, we obtain
\begin{align}
\var\left((\h-\bs{l}_T\otimes\bs{\mu})\right)=\sigma^2\J^{-1}(\rh)\Pp(\rh)\J^{'-1}(\rh),
\end{align}
where
\begin{align}\label{4.13}
\Pp(\rh)=
\begin{pmatrix}
\K(\rh)&\0&\hdots&\0&\0\\
\0&\I_n&\hdots&\0&\0\\
\vdots&\vdots&\ddots&\vdots&\vdots\\
\0&\0&\hdots&\0&\I_n
\end{pmatrix}.
\end{align}
Let $\Om=\sigma^2\J^{-1}(\rh)\Pp(\rh)\J^{'-1}(\rh)$. Then, the distribution of $\h$ is
\begin{align}\label{4.14}
\h|\rh,\bs{\mu},\sigma^2\sim N(\bs{l}_T\otimes\bs{\mu},\,\Om).
\end{align}
Note that when $\bs{\rho}=\mf{0}$, $\bs{\Omega}$ reduces to $\sigma^2\mf{I}_{nT}$, and thus the result in \eqref{4.14} becomes $\h|\bs{\mu},\sigma^2\sim N(\bs{l}_T\otimes\bs{\mu},\,\sigma^2\mf{I}_{nT})$. Consider the following partition of $\bs{\Omega}$:
\begin{align}
\bs{\Omega}=
\begin{pmatrix}
\bs{\Omega}_{11}&\bs{\Omega}_{12}&\hdots&\bs{\Omega}_{1,T-1}&\bs{\Omega}_{1T}\\
\bs{\Omega}_{21}&\bs{\Omega}_{22}&\hdots&\bs{\Omega}_{2,T-1}&\bs{\Omega}_{2T}\\
\vdots&\vdots&\ddots&\vdots&\vdots\\
\bs{\Omega}_{T1}&\bs{\Omega}_{T2}&\hdots&\bs{\Omega}_{T,T-1}&\bs{\Omega}_{TT}\\
\end{pmatrix},
\end{align}
where each $\bs{\Omega}_{st}$ for $s,t=1,2\hdots,T$ is an $n\times n$ sub-matrix of $\bs{\Omega}$. Let $\Omega_{ij,st}$ be the $(i,j)$th element of $\bs{\Omega}_{st}$ for $i,j=1,2,\hdots,n$. Let $r\in\mb{N}$ be a natural even number. Then, the even moments of $y_{it}$ can be expressed as 
\begin{align}
\E\left(y^r_{it}\right)=\E\left(e^{\frac{r}{2}h_{it}}\right)\E(v^r_{it})=\exp\left(\frac{\mu_ir}{2}+\frac{r^2}{8}\Omega_{ii,tt}\right)\gamma(r)
\end{align}
where $\gamma(r)=\frac{r!}{2^{r/2}(r/2)!}$. Then, it follows that $\E(y^4_{it})/\left(\E(y^2_{it})\right)^2-3=3\left(\exp(\Omega_{ii,tt})-1\right)>0$. Thus, our specification suggests that $y_{it}$ has a leptokurtic symmetric distribution. Next, we consider the covariance between $y^r_{it}$ and $y^r_{js}$:
\begin{align}
\text{Cov}\left(y^r_{it},y^r_{js}\right)&=\E\left(e^{\frac{r}{2}(h_{it}+h_{js})}v^r_{it}v^r_{js}\right)-\E\left(e^{\frac{r}{2}h_{it}}v^r_{it}\right)\E\left(e^{\frac{r}{2}h_{js}}v^r_{js}\right)\nonumber\\
&=\gamma^2(r)\exp\left(\frac{r(\mu_i+\mu_j)}{2}+\frac{r^2}{8}\left(\Omega_{ii,tt}+\Omega_{jj,ss}+2\Omega_{ij,ts}\right)\right)\nonumber\\
&-\gamma^2(r)\exp\left(\frac{\mu_ir}{2}+\frac{r^2}{8}\Omega_{ii,tt}\right)\exp\left(\frac{\mu_jr}{2}+\frac{r^2}{8}\Omega_{jj,ss}\right)\nonumber\\
&=\gamma^2(r)\exp\left(\frac{r(\mu_i+\mu_j)}{2}+\frac{r^2}{8}\left(\Omega_{ii,tt}+\Omega_{jj,ss}\right)\right)\left(\exp\left(\frac{r^2}{4}\Omega_{ij,ts}\right)-1\right).
\end{align}
This result indicates that our specification introduces spatial dependence in the outcome variable, since $\text{Cov}\left(y^r_{it},y^r_{js}\right)\ne0$ in general. Note that $\text{Cov}\left(y^r_{it},y^r_{js}\right)=0$ when $\bs{\rho}=\mf{0}$ holds, because  $\left(\exp\left(\frac{r^2}{4}\Omega_{ij,ts}\right)-1\right)=0$.

\section{Posterior Analysis}\label{post}

To introduce a Bayesian MCMC estimation approach, we first transform our model such that the resulting outcome equation is linear in $\h_t$. We then determine the conditional likelihood function of the transformed model by approximating to the distribution of transformed disturbance term with  a Gaussian mixture distribution \citep{Kim:1998, Chib:2002, Omari:2007}. The conditional likelihood function of the transformed model facilitates the sampling steps for $\h_t$ and the auxiliary mixture component indicator defined subsequently. We also provide the conditional likelihood function of the original model, which we use to determine the sampling steps of other parameters in our model. 

We square both sides of \eqref{4.1} and then take the logarithm to obtain 
\begin{align}\label{4.2}
y^*_{it}=h_{it}+v^{*}_{it},
\end{align}
where $y^*_{it}=\log y^2_{it}$ and $v^*_{it}=\log v^2_{it}$. The density of $v^*_{it}$ is highly skewed with a long tail on the left and can be expressed as
\begin{align}\label{2.9} 
p(v^*_{it})=\frac{1}{\sqrt{2\pi}}\exp\left(-\frac{1}{2}(e^{v^*_{it}}-v^*_{it})\right),\quad-\infty<v^*_{it}<\infty,\quad i=1,2,\hdots,n,\,t=1,\hdots,n.
\end{align}
It can be shown that $\E(v^*_{it})\approx-1.2704$ and $\text{Var}(v^*_{it})=\pi^2/2\approx4.9348$. 
Define $\Y^*_t=(y^*_{1t},y^*_{2t},\hdots,y^*_{nt})^{'}$ and  $\V^*_t=(v^*_{1t},v^*_{2t},\hdots,v^*_{nt})^{'}$. Then, in vector form, we have 
\begin{align}\label{4.4}
\Y^*_{t}=\h_{t}+\V^{*}_{t},
\end{align}
In order to convert \eqref{4.4} into a linear Gaussian state-space model, we approximate $p(v^*_{it})$ with an $m$-component Gaussian mixture distribution:
\begin{align}\label{3.1}
p(v^*_{it})\approx\sum_{j=1}^mp_j\times \phi(v^*_{it}|\mu_j,\,\sigma^2_j),
\end{align}
where $\phi(v^*_{it}|\mu_j,\,\sigma^2_j)$ denotes the Gaussian density function with mean $\mu_j$ and variance $\sigma^2_j$,  $p_j$ is the probability of $j$th mixture component and $m$ is the number of components. In particular, we use the ten-component Gaussian mixture distribution suggested by \citet{Omari:2007} to approximate $p(v^*_{it})$. We provide the parameter values of the ten-component Gaussian mixture distribution in Table~\ref{table1}. The parameters in this table are chosen by matching the first four moments of the ten component Gaussian mixture distribution with that of $p(v^{*}_{it})$. This approach has two advantages. First, the Gaussian mixture distribution with the pre-determined parameter values in Table~\ref{table1} provides a well enough approximation to $p(v^*_{it})$  \citep{Omari:2007}. Second, this approach does not pose any estimation difficulties since the mixture parameters in Table~\ref{table1} are pre-determined. 

We can equivalently write \eqref{3.1} in terms of an auxiliary discrete random variable $z_{it}\in\{1,2,\hdots,m\}$ that serves as the mixture component indicator:
\begin{align}\label{4.6}
v^*_{it}|(z_{it}=j)\sim N(\mu_j,\,\sigma^2_j),\quad\text{and}\quad \mathbb{P}(z_{it}=j)=p_j,\quad j=1,2,\hdots,m,
\end{align}
where $\mathbb{P}(z_{it}=j)=p_j$ is the probability that $z_{it}$ takes the $j$th value.  Let $\mathbf{Z}_t=(z_{1t},\hdots,z_{nt})^{'}$, $\mathbf{d}_t=(\mu_{z_{1t}},\hdots,\mu_{z_{nt}})^{'}$ and $\boldsymbol{\Sigma}_t=\Diag(\sigma^2_{z_{1t}},\hdots,\sigma^2_{z_{nt}})$. Then, from \eqref{4.6}, we have $\V^*_t|\mf{Z}_t\sim N(\mathbf{d}_t,\,\boldsymbol{\Sigma}_t)$, which indicates that our model in \eqref{4.4} is now conditionally linear Gaussian given the component indicator variable. Thus, from \eqref{4.4}, we have
\begin{align}\label{4.7}
&\Y^*_t|\mf{Z}_t,\,\h_t\sim N\left(\h_t+\mathbf{d}_t,\,\boldsymbol{\Sigma}_t\right),
\end{align}
which facilitates the sampling steps for $\h_t$ and $\mf{Z}_t$ in our suggested Gibbs sampler given in Algorithm 1. The sampling steps for the remaining parameters requires the following conditional distribution: 
\begin{align}\label{4.8}
\Y_t|\h_t\sim N\left(\mf{0},\,\Hh_t\right).
\end{align}

\definecolor{LightCyan}{rgb}{0.88,1,1}
\begin{table}
\begin{center}
\caption{The ten-component Gaussian mixture for $p(v^*_{it})$ } 
\label{table1} 
\setlength{\tabcolsep}{5pt} 
\renewcommand{\arraystretch}{1} 
\begin{tabular*}{0.65\textwidth}{@{\extracolsep{\fill} }ccrc} 
\hline\hline
Components& $p_j$ &\multicolumn{1}{c}{$\mu_j$}&$\sigma^2_j$ \\
\hline
1 &0.00609 &1.92677& 0.11265 \\ 
2&0.04775& 1.34744& 0.17788\\
3 &0.13057& 0.73504& 0.26768\\
4&0.20674 &0.02266& 0.40611\\
5 &0.22715&-0.85173 &0.62699\\
6&0.18842&-1.97278& 0.98583\\
7&0.12047&-3.46788& 1.57469\\
8 &0.05591&-5.55246& 2.54498\\
9&0.01575 &-8.68384 &4.16591\\
10&0.00115&-14.65000&7.33342\\
\hline\hline
\end{tabular*}
\end{center}
\end{table}
We are now in a position to design a Gibbs sampler by using our results on (i) the mixture component indicators in \eqref{4.6}, (ii) the conditional likelihood function of transformed model in \eqref{4.7}, (iii) the conditional likelihood function of $\Y_t$ in \eqref{4.8} and (iv) the distribution of $\h$ in \eqref{4.14}. Let $\Y=(\Y^{'}_1,\hdots,\Y^{'}_T)^{'}$ and $\Z=(\Z^{'}_1,\hdots,\Z^{'}_T)^{'}$. The joint posterior distribution  $p(\mathbf{h},\mathbf{Z},\bs{\mu},\rh,\sigma^2|\mathbf{Y})$ can be expressed as
\begin{align}\label{4.15}
 p(\mathbf{h},\mathbf{Z},\bs{\mu},\rh,\sigma^2|\mathbf{Y})&\propto p(\Y^{*}|\Z, \h)\times p(\h|\rh,\bs{\mu},\sigma^2)\times p(\Z)\times  p(\bs{\mu})\times p(\rh)\times p(\sigma^2).
\end{align}
Then, our suggested Gibbs sampler for generating draws from $p(\mathbf{h},\mathbf{Z},\bs{\mu},\rh,\sigma^2|\mathbf{Y})$ consists of the steps given in Algorithm 1.
\begin{algorithm}[Estimation Algorithm]\label{a1}
\leavevmode   \normalfont
\begin{enumerate}
\item Sampling step for $\Z$:  Note that $\Z_t$ is a discrete random variable, and its conditional posterior probability mass function is
\begin{align}
p(\Z_t|\Y_t,\h_t,\bs{\mu},\rh,\sigma^2)\propto p(\Z_t)p(\Y^{*}_t|\Z_t,\h_t)=\prod_{i=1}^np(y^{*}_{it}|z_{it},h_{it})p(z_{it}),
\end{align}
for $t=1,\hdots,T$. Thus,
\begin{align}\label{4.17}
\mathbb{P}(z_{it}=j|y^{*}_{it})=\frac{p_j\phi(y^{*}_{it}|h_{it}+\mu_j,\,\sigma^2_j)}{\sum_{k=1}^{10}p_k\phi(y^{*}_{it}|h_{it}+\mu_k,\,\sigma^2_k)},\,j=1,\hdots,10,\,i=1,\hdots,n,
\end{align} 
for $t=1,\hdots,T$, where the denominator is the normalization constant.
\item Sampling step for $\h$: Let $\Si=\Diag\left(\Si_1,\hdots,\Si_T\right)$ and $\mf{d}=(\mf{d}^{'}_1,\hdots,\mf{d}^{'}_T)^{'}$. Using standard regression results on $p(\h|\Y,\Z,\bs{\mu},\rh,\sigma^2)\propto \prod_{t=1}^Tp(\Y^{*}_t|\Z,\h_t)p(\h|\rh,\bs{\mu},\sigma^2)$, we obtain
\begin{align}
\h|\Y,\Z,\bs{\mu},\rh,\sigma^2\sim N(\hat{\mf{b}}_h,\hat{\B}_h),
\end{align}
where 
\begin{align*}
\hat{\B}_h=\left(\Om^{-1}+\Si^{-1}\right)^{-1},\quad\hat{\bs{b}}_h=\hat{\B}_h\left(\Om^{-1}(\bs{l}_T\otimes\bs{\mu})+\Si^{-1}(\Y^{*}-\mf{d})\right).
\end{align*}
\item Sampling step for $\bs{\mu}$: Using \eqref{4.9} and \eqref{4.13}, we can find that
\begin{align*}
\Om^{-1}=
\begin{pmatrix}
\Om^{*}_{11} & \Om^{*}_{12} &\0& \ldots&\0& \0 \\
\Om^{*}_{21}&\Om^{*}_{22}&\Om^{*}_{23}&\hdots&\0&\0\\
\vdots & \ddots &\ddots &\ddots&\vdots& \vdots \\
\0&\0& \0 &\ldots&\Om^{*}_{T-1,T-1}&\Om^{*}_{T-1, T}\\
\0&\0& \0 &\ldots&\Om^{*}_{T,T-1}&\Om^{*}_{TT}
\end{pmatrix},
\end{align*}
where
\begin{align*}
&\Om^{*}_{11}= \sigma^{-2}\Ss^{'}(\rho_1)\K^{-1}(\rh)\Ss(\rho_1)+\sigma^{-2}\A^{'}(\rho_2,\rho_3)\A(\rho_2,\rho_3),\quad \Om^{*}_{TT}=\sigma^{-2}\Ss^{'}(\rho_1)\Ss(\rho_1),\\
&\Om^{*}_{ii}=\sigma^{-2}\Ss^{'}(\rho_1)\Ss(\rho_1)+\sigma^{-2}\A^{'}(\rho_2,\rho_3)\A(\rho_2,\rho_3),\quad i=2,\hdots,T-1,\\
&\Om^{*}_{i,i+1}=\Om^{*'}_{i+1,i}=-\sigma^{-2}\A^{'}(\rho_2,\rho_3)\Ss(\rho_1),\quad i=1,\hdots,T-1.
\end{align*}
Then, from $p(\bs{\mu}|\Y,\h,\Z,\bs{\mu},\rh,\sigma^2)\propto p(\h|\rh,\bs{\mu},\sigma^2)p(\bs{\mu})$, we obtain
\begin{align}
\bs{\mu}|\Y,\h,\Z,\bs{\mu},\rh,\sigma^2\sim N(\hat{\mf{b}}_{\mu},\hat{\B}_{\mu}),
\end{align}
where
$$
\hat{\B}_{\mu}=\left(\B^{-1}_{\mu}+\sum_{j=1}^T\sum_{i=1}^T\Om^{*}_{ij}\right)^{-1},\quad
\hat{\mf{b}}_{\mu}=\hat{\B}_{\mu}\left(\B^{-1}_\mu\bs{b}_\mu+\sum_{j=1}^T\sum_{i=1}^T\Om^{*}_{ij}\h_i\right)
$$
\item Sampling step for $\sigma^2$: From $p(\sigma^2|\Y,\h,\bs{\mu},\rh)\propto p(\h|\rh,\bs{\mu},\sigma^2)p(\sigma^2)$, we obtain
\begin{align*}
\sigma^2|\Y,\h,\Z,\bs{\mu},\rh\sim\text{IG}(\hat{a},\,\hat{b}),
\end{align*}
where 
$$
\hat{a}=a+nT/2,\quad\hat{b}=b+\frac{1}{2}\left(\h-(\bs{l}_T\otimes\bs{\mu})\right)^{'}\left(\J^{'}(\rh)\Pp^{-1}(\rh,\la)\J^{}(\rh)\right)\left(\h-(\bs{l}_T\otimes\bs{\mu})\right).
$$
\item Sampling step for $\rh$: The conditional posterior density of $\rh$ does not take any known form in our model. We use the adaptive Metropolis (AM) algorithm suggested in \citet{Haario:2001} and \citet{Roberts:2009} to generate draws from $p(\rh|\Y,\h,\Z,\sigma^2)$.\footnote{\citet{Han:2016} and \citet{Han:2017} use this algorithm to generate draws for the spatial parameters in spatial panel data models. Their results show that this algorithm can perform satisfactorily.}   At the iteration $g$, we use the following proposal distribution to generate the candidate value $\tilde{\bs{\rho}}$: 
\begin{align*}
f_g\left(\rh|\rh^{(0)},\hdots, \rh^{(g-1)}\right)=
\begin{cases}
N\left(\rh^{(g-1)},\,\frac{(0.1)^2}{3}\times \I_3\right),\quad\text{for}\quad g\leq g_0,\\
0.95\times N\left(\rh^{(g-1)},\,\frac{c(2.38)^2}{3}\times\text{Cov}\left(\rh^{(0)},\hdots, \rh^{(g-1)}\right)\right)\\
\quad\quad +0.05\times N\left(\rh^{(g-1)},\,\frac{(0.1)^2}{3}\times \I_3\right),\quad\text{for}\quad g>g_0,
\end{cases}
\end{align*}
where $g_0$ is the length of initial sampling period, $\text{Cov}\left(\rh^{(0)},\hdots, \rh^{(g-1)}\right)$ is the empirical covariance matrix of historical draws given by $\text{Cov}\left(\rh^{(0)},\hdots, \rh^{(g-1)}\right)=\frac{1}{g}\sum_{j=0}^{g-1}\bs{\rho}^{(j)}\bs{\rho}^{(j)'}-\bar{\bs{\rho}}^{(g-1)}\bar{\bs{\rho}}^{(g-1)'}$ with $\bar{\bs{\rho}}^{(g-1)}=\frac{1}{g}\sum_{j=0}^{g-1}\bs{\rho}^{(j)}$, and $c$ is a scalar tuning parameter used to achieve a reasonable acceptance rate. We then check whether $\tilde{\rh}$ satisfies the stability conditions in
Section \ref{spec}. If not, we regenerate $\tilde{\rh}$ until it meets the stability conditions. We compute the following acceptance probability:
\begin{align*}
\mathbb{P}(\rh^{(g-1)}, \tilde{\rh})=\min\left(\frac{p\left(\h|\tilde{\rh},\bs{\mu}^{(g)},\sigma^{2(g)}\right)}{p\left(\h|\rh^{(g-1)},\bs{\mu}^{(g)}, \sigma^{2(g)}\right)},\,1\right),
\end{align*}
where $p(\h|\rh,\bs{\mu},\sigma^{2})$ is given in \eqref{4.14}. Finally, we return $\tilde{\rh}$ with probability $\mathbb{P}(\rh^{(g-1)}, \tilde{\rh})$; otherwise return $\rh^{(g-1)}$.
\end{enumerate}
\end{algorithm}
\begin{rmk}
In Step 1, the result in \eqref{4.17} indicates that the mixture components are conditionally independent given $y^{*}_{it}$. Thus, each component is a discrete random variable taking integer values in the interval $[1,10]$ with the conditional posterior probability $\mathbb{P}(z_{it}=j|y^{*}_{it})$. The conditional posterior results in Steps 2, 3 and 4 are obtained from a standard Bayesian analysis as in a linear regression model.  In the AM algorithm described in Step 5, the proposal distribution $f_g\left(\rh|\rh^{(0)},\hdots, \rh^{(g-1)}\right)$ has two parts. The first part is  $N\left(\rh^{(g-1)},\,\frac{(0.1)^2}{3}\times \I_3\right)$, and is used when the number of iterations is less than or equal to $g_0$. The second part consists of two normal distributions. The first component is specified as  $N\left(\rh^{(g-1)},\,\frac{c(2.38)^2}{3}\times\text{Cov}\left(\rh^{(0)},\hdots, \rh^{(g-1)}\right)\right)$, where the covariance matrix is determined from the historical MCMC draws of $\rh$. The second component is $N\left(\rh^{(g-1)},\,\frac{(0.1)^2}{3}\times \I_3\right)$. The candidate values generated  from $f_g\left(\rh|\rh^{(0)},\hdots, \rh^{(g-1)}\right)$ are subject to the stability conditions given in \eqref{3.5}. Finally, we adjust the tuning parameter $c$ during the estimation to achieve an acceptance rate that falls between 40 percent and 60 percent.
\end{rmk}
\begin{rmk}
In the sampling step for $\rh$, $\mathbb{P}(\rh^{(g-1)}, \tilde{\rh})$ is calculated at each pass of the sampler, and therefore, $p(\h|\rh,\bs{\mu},\sigma^2)$ is evaluated twice at each pass of the sampler. In other words, $p(\h|\rh,\bs{\mu},\sigma^2) = (2\pi)^{-nT/2}|\Om|^{-1/2}\exp\left(-\frac{1}{2}(\h - \bs{l}_T\otimes\bs{\mu})^{'}\Om^{-1}(\h - \bs{l}_T\otimes\bs{\mu})\right)$ must be calculated twice. Since $\Om=\sigma^2\J^{-1}(\rh)\Pp(\rh)\J^{'-1}(\rh)$, we have $|\Om|^{-1/2} =  (\sigma^2)^{-nT/2}|\J(\rh)||\Pp(\rh)|^{-1/2}$. From \eqref{4.9}, since $\J(\rh)$ is a triangular matrix, we have $|\J(\rh)|=\prod_{t=1}^T|\Ss(\rho_1)|=|\Ss(\rho_1)|^T$. Also, from \eqref{4.13}, since $|\Pp(\rh)|$ is a block-diagonal matrix, we have $|\Pp(\rh)|=|\K(\rh)|$. 
\end{rmk}

\section{Simulations}\label{sim}

In this section, we provide simulation evidence to assess sampling properties of the suggested Bayesian algorithm. The data generating process follows \eqref{2.1} and \eqref{2.2}. More specifically, the elements of $\V_{t}$ and $\U_t$ are drawn independently from the standard normal distribution for $t=1,2,\hdots,T$. Therefore, the value of $\sigma^2$ is set to $0.25$ in all experiments. The elements of $\bs{\mu}$ are drawn independently from the normal distribution with mean $3.3$ and standard deviation $0.35$. To initialize the process, we use \eqref{4.14}, and the series expression for $\K(\rh)$ is truncated at 15. We consider two sets of values for $\rh$, $\{(0.6,0.35,-0.025),(0.3,0.65,-0.025)\}$. These parameter values are chosen to ensure that the data generating process mimics the findings from our empirical application in the next section. The number of spatial units $n$ is set to $98$, and the number of time periods $T$ is fixed at $50$. 

For the spatial weights matrix $\W$, we consider row-standardized rook and queen contiguity weights matrices. To this end, we first generate a vector containing a random permutation of the integers from $1$ to $n$ without repeating elements. Then, we reshape this vector into an $k\times m$ rectangular lattice, where $m=n/k$.  In the case of rook contiguity, we set $w_{ij} = 1$ if the $j$th observation is adjacent (left/right/above or below) to the $i$th observation on the  lattice. In the case of queen contiguity, we set $w_{ij}=1$ if the $j$th observation is adjacent to, or shares a border with the $i$th observation. We set $k=7$, and row-normalize all spatial weights matrices. For the prior distributions, we consider the following: $\sigma^2\sim\text{IG}(3,\,2)$ and $\bs{\mu}\sim N(\mf{0},\,10\I_n)$. The length of the Markov chain is $22000$ draws, and the first $2000$ draws are discarded to dissipate the effects of the initial values. 

\begin{figure}[ht!]
    \centering  
     \includegraphics[width=3.1in, height=2in]{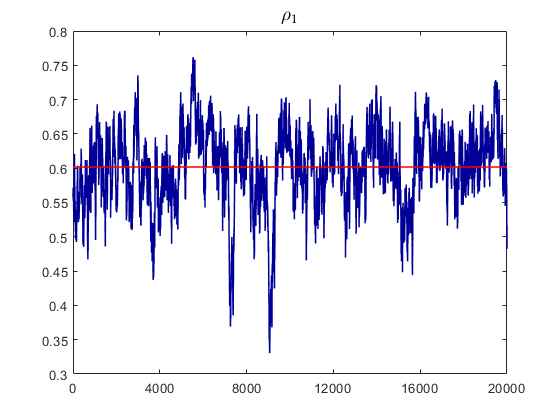}
     \hspace{-0.5cm}
      \includegraphics[width=3.1in, height=2in]{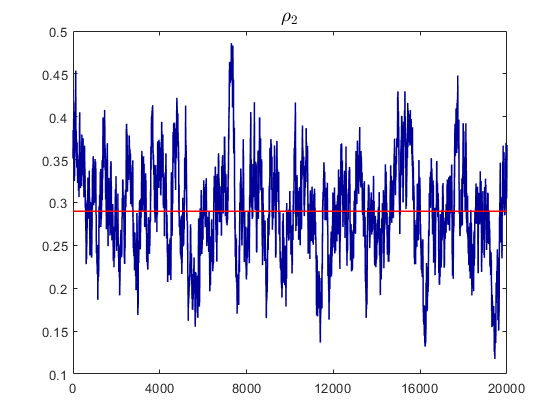}
      \includegraphics[width=3.1in, height=2in]{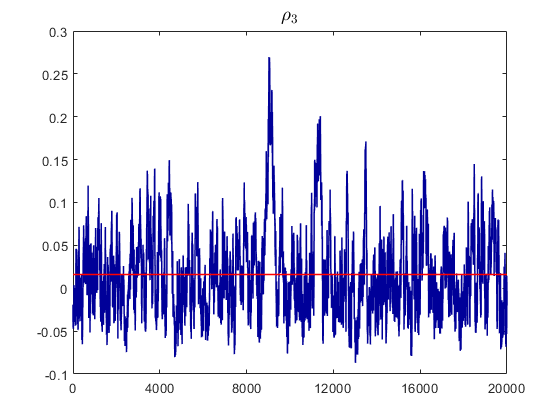}
     \hspace{-0.5cm}
       \includegraphics[width=3.1in, height=2in]{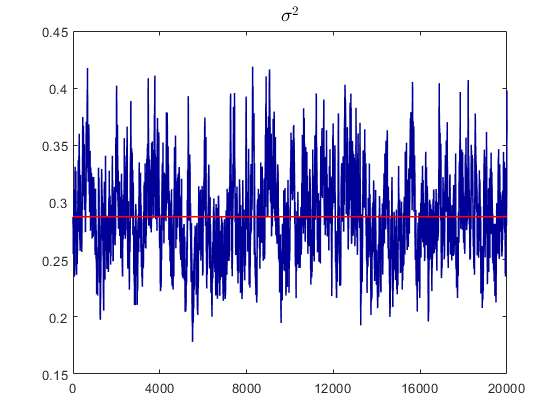}
     \caption{Trace plots for $\rh = (0.6,0.35,-0.025)$ and $\W$ is a queen contiguity matrix}
\label{c1q1}		
\end{figure}

\begin{figure}[ht!]
    \centering  
     \includegraphics[width=3.1in, height=2in]{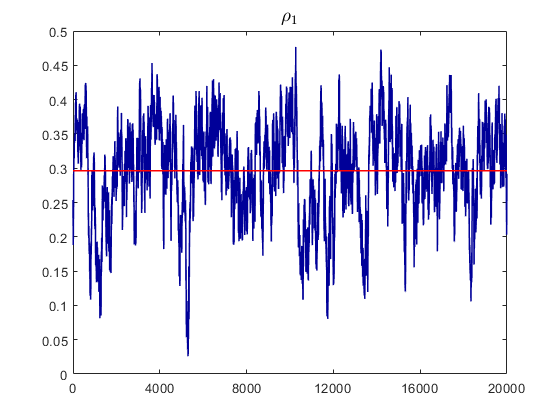}
     \hspace{-0.5cm}
      \includegraphics[width=3.1in, height=2in]{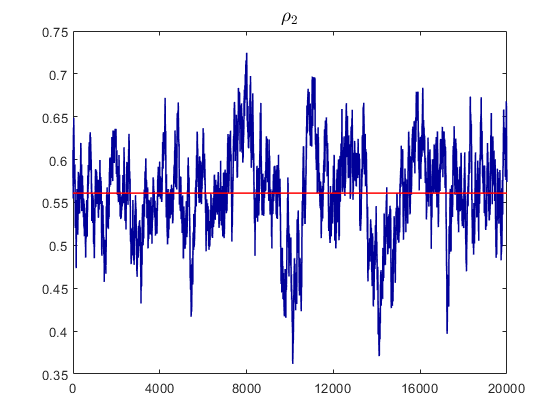}
      \includegraphics[width=3.1in, height=2in]{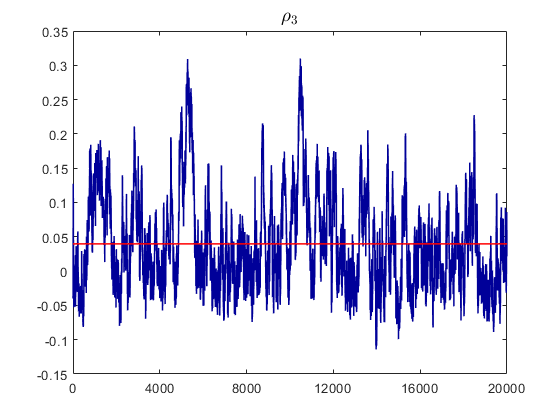}
     \hspace{-0.5cm}
       \includegraphics[width=3.1in, height=2in]{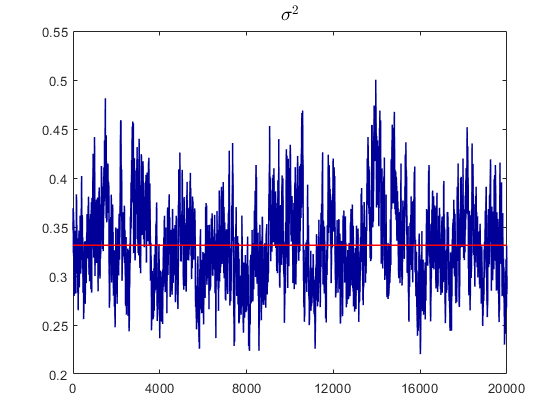}
     \caption{Trace plots for $\rh = (0.30,0.65,-0.025)$ and $\W$ is a queen contiguity matrix}
\label{c2q1}		
\end{figure}

To determine the adequacy of the length the chains and their mixing properties, some exemplary trace plots are provided in Figure~\ref{c1q1} and \ref{c2q1}. For the sake of brevity, we only present the results for the queen contiguity case. In these trace plots, the red solid lines correspond to the estimated posterior means. We observe that the Bayesian estimator performs satisfactorily and seems to mix well in all cases. Note also that for both $\rh$ and $\sigma^2$, the $95\%$ credible intervals contain the true values chosen in the experiments.

\begin{figure}[ht!]
    \centering  
     \includegraphics[width=3.4in, height=2.5in]{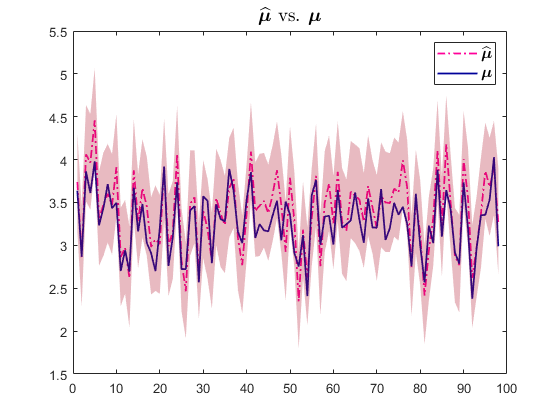}
     \hspace{-1cm}
      \includegraphics[width=3.4in, height=2.5in]{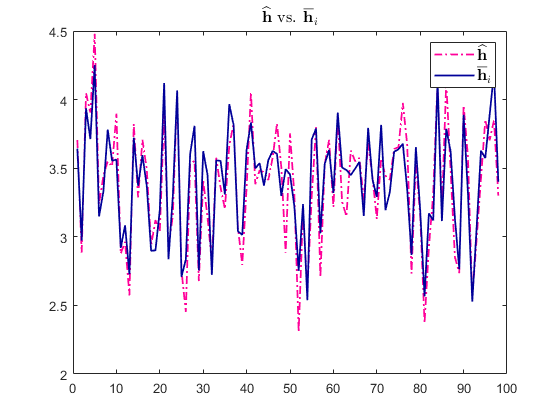}
      \includegraphics[width=3.4in, height=2.5in]{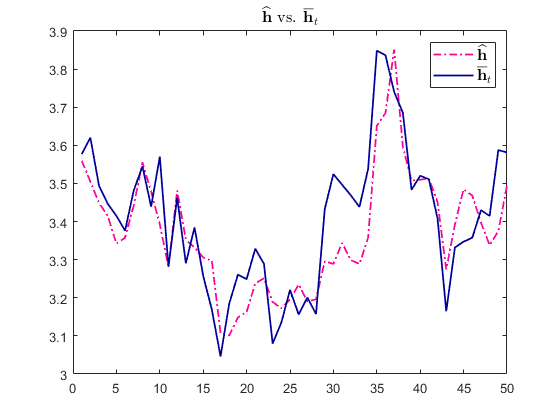}
     \caption{Estimates of $\bs{\mu}$ and $\h$ (dashed lines) and their corresponding true data-generating values (solid lines) for $\rh = (0.6,0.35,-0.025)$ and $\W$ is a queen contiguity matrix}
\label{c1q2}		
\end{figure}

\begin{figure}[ht!]
    \centering  
     \includegraphics[width=3.4in, height=2.5in]{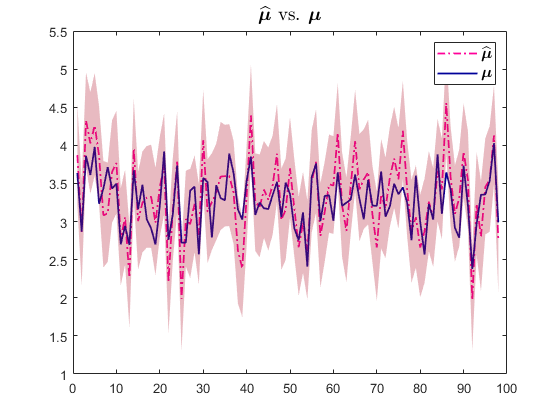}
     \hspace{-1cm}
      \includegraphics[width=3.4in, height=2.5in]{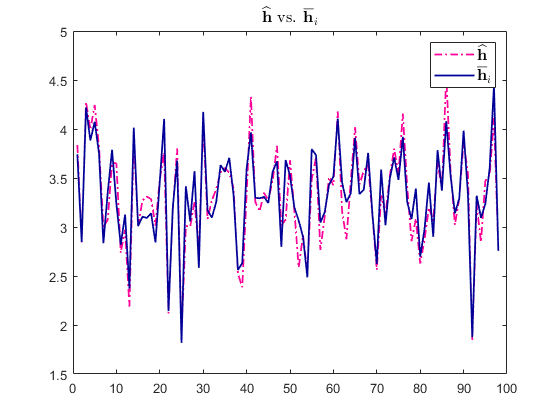}
      \includegraphics[width=3.4in, height=2.5in]{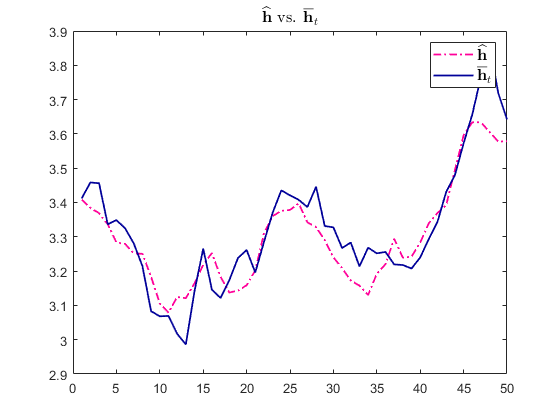}
     \caption{Estimates of $\bs{\mu}$ and $\h$ (dashed lines) and their corresponding true data-generating values (solid lines) for $\rh = (0.30,0.65,-0.025)$ and $\W$ is a queen contiguity matrix}
\label{c2q2}		
\end{figure}

For the $n\times 1$ vector $\bs{\mu}$ and the $nT\times 1$ vector $\h$, we provide evidence on the performance of our Bayesian estimator in Figure~\ref{c1q2} and \ref{c2q2}. In these plots, the true values are represented with solid lines and the estimates are presented with dashed lines. In the first panel, we observe that the estimated posterior means for the components of $\bs{\mu}$ are in general close to the true values. Here, the shaded region refers to the $95\%$ credible interval. For $\h$, we calculate the average of true values over $n$ and over $T$ respectively, and plot them against the average posterior means over $n$ and over $T$, respectively. The second panel presents the case where the average is taken over $n$, and the last panel is the case where the average is taken over $T$. We observe that the Bayesian estimator performs satisfactorily in terms of capturing the log-volatility over cross-sections as well as over time.



\section{Air Quality Modeling and Environmental Risks}\label{emp}

In this section, we demonstrate the usage of the dynamic stochastic volatility model using an empirical example from environmental science. Like for applications in financial economics, the volatility of the process can be interpreted as risks (i.e., environmental risks in our case). Previous ecological studies mostly focused on changes in the mean behavior \citep[cf.][]{wilby2009review}, but also an increased variation might be harmful, e.g., \citet{vasseur2014increased} showed that an increased temperature variation (i.e., changes in the variance) poses a greater risk than global warming (i.e., changes in the mean). Similar results on this topic were found by \citet{paaijmans2013temperature} and \citet{tewksbury2008putting}. Moreover, there is also a connection between climate changes and the financial market, such that climate variations might have an impact on the risk of financial markets (see also \citealt{giglio2021climate,hong2020climate}).

\begin{figure}
  \includegraphics[width = 0.5\textwidth]{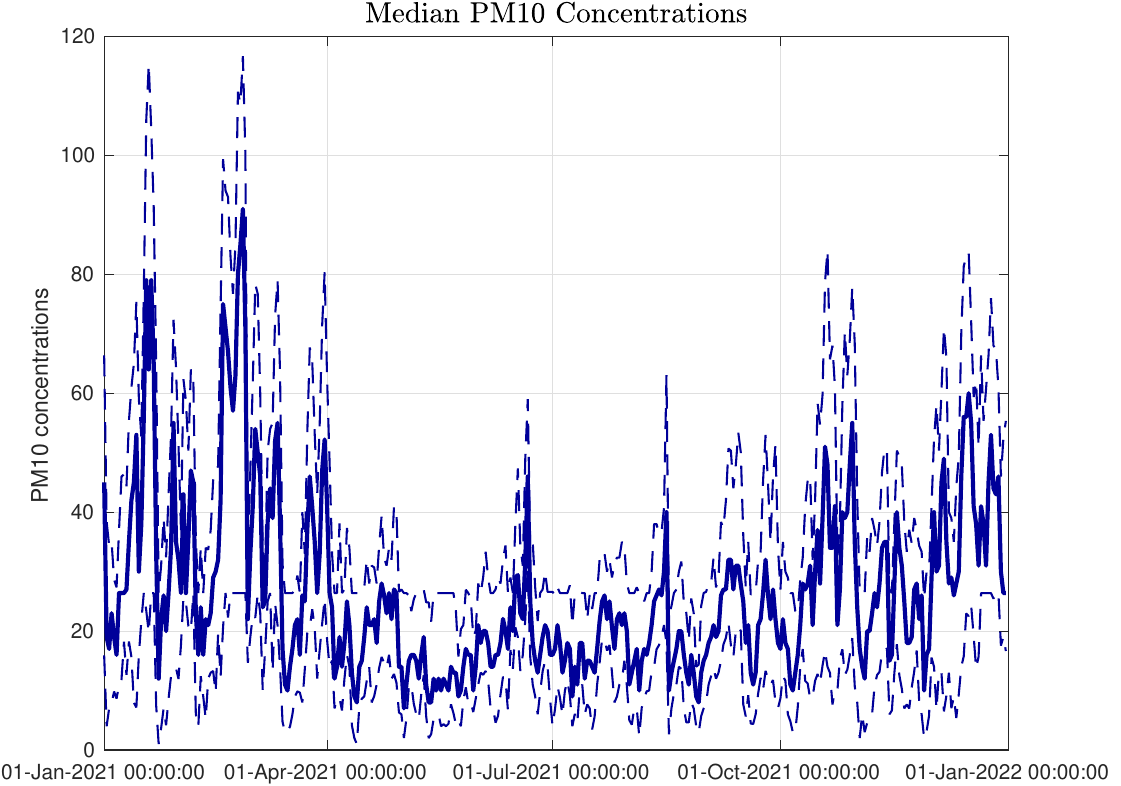}
  \includegraphics[width = 0.5\textwidth]{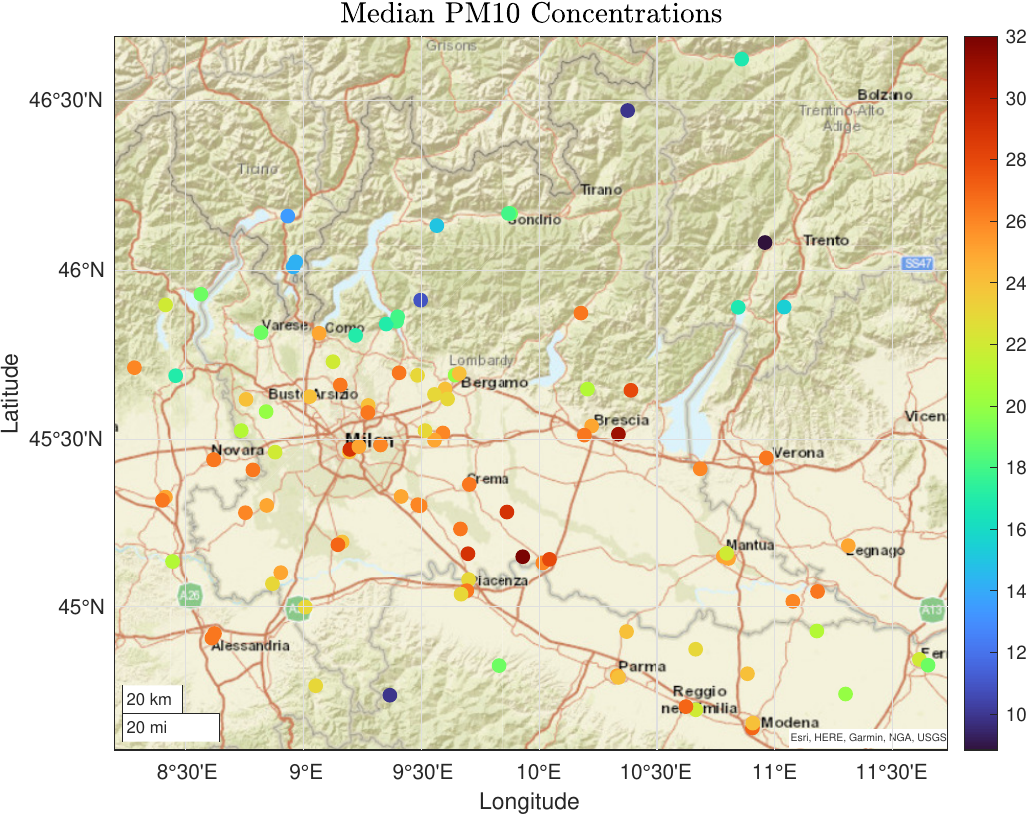}
  \caption{Median $PM_{10}$ concentrations ($\mu g / m^{3}$) and the 5\% and 95\% quantiles across space displayed as time series (left) and median concentrations across time shown on a map (right).}\label{fig:medians}
\end{figure}

 In this article, we analyze the variation of fine dust concentrations of particles having a diameter less than 10 $\mu m$, $PM_{10}$, in Lombardy, Northern Italy. The region is surrounded by the Alps from the west, so that the wind circulation is reduced and Lombardy becomes one of the regions in Europe with the lowest air quality \citep[see][]{fasso2022spatiotemporal}. In the following empirical analysis, we use the daily $PM_{10}$ concentrations from 1.1.2021 to 31.12.2021 from the official monitoring stations of the regional environmental authority, ARPA Lombardia \citep{maranzano2022air}. The data are open-source provided by the Agrimonia project \citep{zenodo_agrimonia,agrimonia_description}. In total, there are $n = 103$ measurement stations and $T = 365$ daily observations. To provide a first overview of the dataset, we depicted the median concentrations in $\mu g / m^{3}$ across space and time as a time-series plot (Figure \ref{fig:medians}, left) and displayed on map (Figure \ref{fig:medians}, right), respectively. 

For our analysis, we first estimated a spatial panel model to describe the mean variations. The spatial correlations have been modeled in an autoregressive manner. To be precise, the mean model is given by
\begin{equation}
	\mathbf{C}_t = \psi \mathbf{W} \mathbf{C}_t  + \mathbf{X}_t \boldsymbol{\beta} + \mathbf{W} \mathbf{X}_t \boldsymbol{\gamma} + \mathbf{s} + \mathbf{E}_t \, .
\end{equation} 
The outcome variable $\mathbf{C}_t$ is the $n$-dimensional vector of $PM_{10}$ concentrations at time point $t$, $\mathbf{W}$ is spatial weights matrix which we specify below in more detail, $\mathbf{X}_t$ is an $n\times p$ matrix of exogenous regressors, and $\mathbf{E}_t$ denotes the model error terms. Spatial interactions are included via the spatial autoregressive term $\psi \mathbf{W} \mathbf{C}_t$ with an unknown autoregressive parameter $\psi$. Moreover, spatial fixed effects $\mathbf{s}$ are present for each station (even if non-significant), because there are different types of stations included in the data set, e.g., urban traffic stations located at major roads in the cities, or rural background stations located in the Alps. The spatial fixed effects describe the station-specific $PM_{10}$ concentrations and thereby serve as the model intercept. We refer to this model as Model A. Moreover, we consider the same model with spatial and temporal fixed effects as an alternative model B, that is,
\begin{equation}
	\mathbf{C}_t = \psi \mathbf{W} \mathbf{C}_t  + \mathbf{X}_t \boldsymbol{\beta} + \mathbf{W} \mathbf{X}_t \boldsymbol{\gamma} + \mathbf{s} +  a_t\mathbf{1}_n + \mathbf{E}_t \, .
\end{equation} 
The temporal fixed effects $a_t\mathbf{1}_n$ will remove any additional (station invariant) seasonal variation across time, which is not explained by the exogenous regressors. More precisely, we included $p = 4$ covariates that are known to influence PM concentrations, namely the maximum height of the planetary boundary layer (PBL), the relative humidity, the air temperature, and the pressure level. All covariates are available as daily observations for each measurement station. Moreover, they were standardized to compare the size of the effects. The spatial weights matrix has been chosen as row-standardized binary contiguity matrix, where all locations within 15 miles are considered as adjacent stations. On average, each stations has 5.2621 neighbors leading to a sparsity level of $\mathbf{W}$ of 94.95 \%. We estimated all parameters $\{\mathbf{s}, \boldsymbol{\beta}, \boldsymbol{\gamma}, \psi, \sigma_{E}^2\}$ using the maximum likelihood approach implemented in the spatial econometrics MATLAB toolbox (cf. \citealt{bivand2015comparing,lesage1999spatial}). The resulting estimates including their asymptotic 95\% confidence intervals are summarized in Table \ref{tab:mean_model}.

\begin{sidewaystable}
  \centering
  \caption{Estimated parameters of the mean model}\label{tab:mean_model}
  \begin{tabular}{ll ccc ccc}
  \hline\hline
    & & \multicolumn{3}{c}{Model A (spatial fixed effects)}                       & \multicolumn{3}{c}{Model B (spatiotemporal fixed effects)}                       \\
    & & Estimate & Standard error & 95\% confidence       & Estimate & Standard error & 95\% confidence       \\
    & &          &                & interval (asymptotic) &          &                & interval (asymptotic) \\
    \hline
    \multicolumn{2}{l}{Average fixed effects}       &         &         &                      &         &         &                      \\
    & Spatial fixed effects                         &  0.0000 & 16.3766 & (-32.0975, 32.0975)  &  0.0000 & 13.4717 & (-26.4040, 26.4040)  \\
    & Temporal fixed effects                        &  /      & /       &  /                   &  0.0000 &  5.8340 & (-11.4343, 11.4343)  \\
    \multicolumn{2}{l}{Regressors}                  &         &         &                      &         &         &                      \\
    & $\beta_1$ (max PBL height)                    & -2.6270 &  0.1989 & (-3.0169, -2.2371)   & -0.7158 &  0.2038 & (-1.1152, -0.3164)   \\
    & $\beta_2$ (temperature)                       & -3.0972 &  0.2205 & (-3.5295, -2.6650)   & -2.8055 &  0.4430 & (-3.6739, -1.9372)   \\
    & $\beta_3$ (relative humidity)                 & -0.2110 &  0.1787 & (-0.5612,  0.1392)   &  0.1272 &  0.1785 & (-0.2226,  0.4771)   \\
    & $\beta_4$ (air pressure)                      & 20.7476 &  1.0724 & (18.6457, 22.8494)   & 16.5707 &  2.0569 & (12.5392, 20.6021)   \\
    \multicolumn{2}{l}{Spatially lagged regressors} &         &         &                      &         &         &                      \\
    & $\gamma_1$ (max PBL height)                   &  1.3812 &  0.2104 & (0.9688, 1.7935)     &  0.6939 &  0.2057 & (0.2908, 1.0970)     \\
    & $\gamma_2$ (temperature)                      &  2.3897 &  0.2270 & (1.9449, 2.8346)     &  1.7405 &  0.2334 & (1.2831, 2.1980)     \\
    & $\gamma_3$ (relative humidity)                &  0.2534 &  0.1925 & (-0.1238, 0.6307)    &  0.2444 &  0.1916 & (-0.1310, 0.6199)    \\
    & $\gamma_4$ (air pressure)                     &-12.9406 &  1.1049 & (-15.1061, -10.7751) & -6.3344 &  1.0464 & (-8.3852, -4.2835)   \\
    \multicolumn{2}{l}{Spatial dependence}          &         &         &                      &         &         &                      \\
    & $\psi$                                        &  0.7240 &  0.0028 & (0.7185, 0.7294)     &  0.5050 &  0.0045 & (0.4962, 0.5138)     \\
    \multicolumn{2}{l}{Error Variance}              &         &         &                      &         &         &                      \\
    & $\sigma_{E}^2$                                & 62.8830 &         &                      & 56.4973 &         &                      \\
    & Coefficient of                                &         &         &                      &         &         &                      \\
    & determination $R^2$                           &  0.7829 &         &                      &  0.8049 &         &                      \\
    \hline
  \end{tabular}
\end{sidewaystable}

\begin{figure}
  \includegraphics[width = 0.5\textwidth]{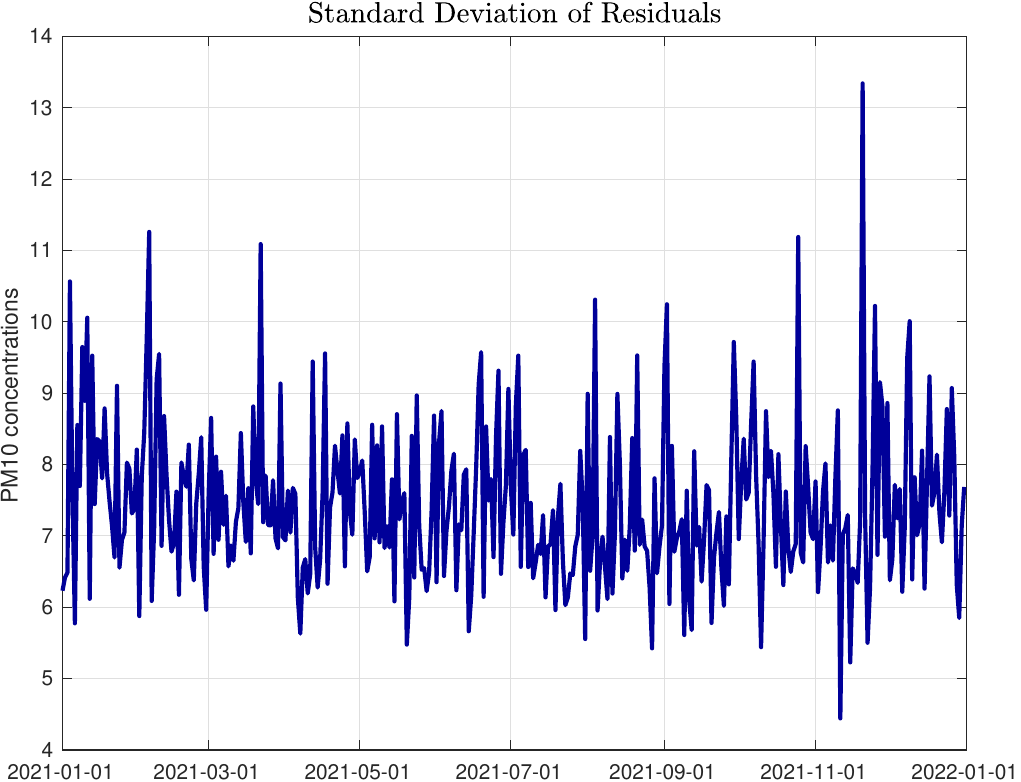}
  \includegraphics[width = 0.5\textwidth]{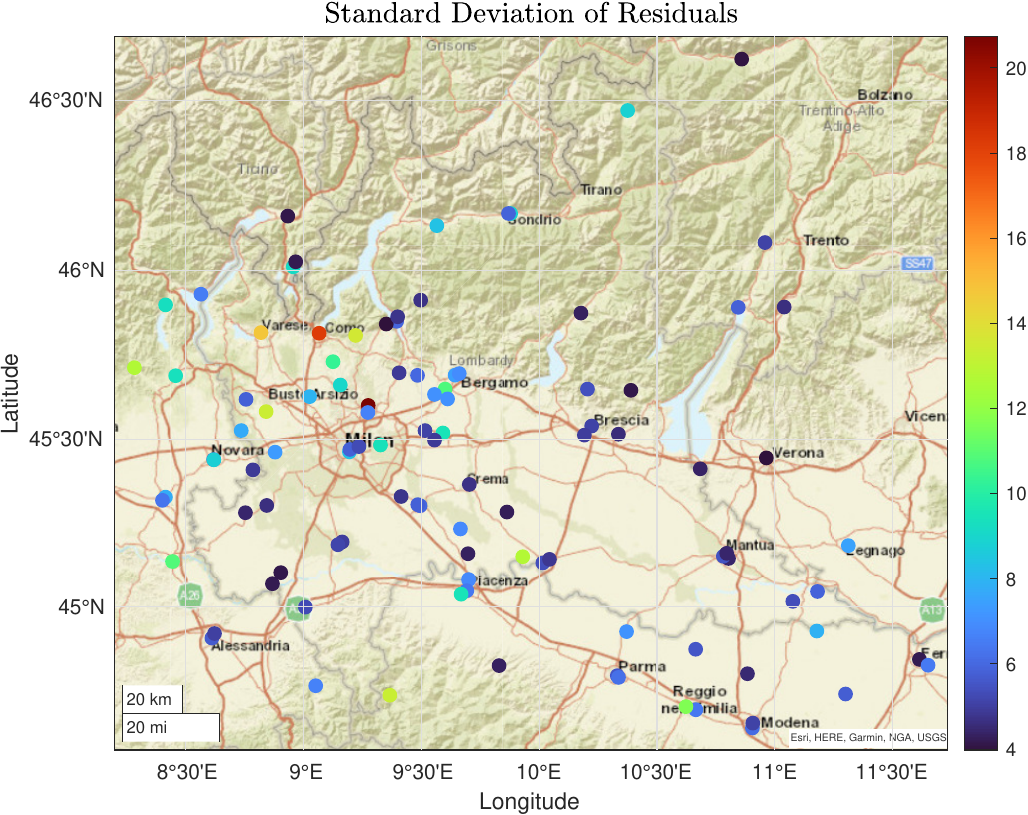}
  \caption{Standard deviation of residuals $\mathbf{E}_t$}\label{fig:residuals}
\end{figure}

The estimated parameters are in line with our expectations and we see a good general fit of the model ($R^2 = 0.7829$ and $0.8049$ with only spatial and spatiotemporal fixed effects, respectively). However, for this paper, our focus is not on the mean variations but on the model errors $\mathbf{E}_t$, i.e., the model uncertainties. Hence, we do not go into further detail on the interpretation of the mean models. The errors of the mean model represent the unexplained variations and, therefore, the environmental risks. When considering the standard deviation of $\mathbf{E}_t$ at each time point and/or for each measurement station, we see that the error variance is varying across space and time (see Figure \ref{fig:residuals}). There are time periods of increased variations, so-called volatility clusters, e.g. at the end of the year. Moreover, we observe a similar clustering across space. Measurement stations with a higher volatility are located in close proximity, e.g., North-East of Milan or around Brescia. This provides motivation for estimating a dynamic stochastic volatility model for $\mathbf{E}_t$ in the second step, that is,
\begin{align}
&\mathbf{E}_{t}=\Hh^{1/2}_t\V_{t}
\end{align}
as defined in \eqref{2.1} and \eqref{2.2}. 

Table \ref{tab:svm_model} reports the median of the posterior draws for each parameter including the corresponding 95\% credible intervals (500,000 posterior draws, and 100,000 burn-in draws). The space-time interaction parameters $\rho_1$, $\rho_2$, and $\rho_3$ represent the uncertainty/risk spillovers. In general, we observe moderate instantaneous spatial interactions (i.e., $\hat{\rho}_1 = 0.5943$ for Model A and $\hat{\rho}_1 = 0.3037$ for Model B) and temporal autoregressive interactions (i.e., $\hat{\rho}_2 = 0.3781$ for Model A and $\hat{\rho}_2 = 0.6699$ for Model B), while spatiotemporal effects appear to be of minor importance (i.e., $\hat{\rho}_3 = -0.0278$ and $\hat{\rho}_3 = -0.0236$ for Models A and B, respectively). The posterior mean estimates for both the spatial and temporal autoregressive parameters are positive. That is, if the environmental risk (log-volatility $h_{it}$) is high it is likely to spillover to the neighboring regions and to future time points. For this reason, spatial and temporal volatility clusters are formed. The negative signs of the posterior mean estimate for the spatiotemporal lag are opposing this behavior, but they are close to zero. Moreover, we observe that the spatial spillovers are dominating the temporal ones for Model A without temporal fixed effects. That is, the missing temporal variation of mean model could be picked up within the uncertainty model. When including both temporal and spatial fixed effects (i.e., Model B), the spatial dependence in the volatility of the errors is reduced, while the temporal dependence (i.e., $\rho_2$) is significantly higher. When looking at the overall volatility level represented by the average $\mathbf{h}_t$, we see that the model uncertainty is reduced because of the additional temporal fixed effects in the mean equation. Nevertheless, all space-time interaction parameters are significantly different from zero indicating a space-time dependence in the local uncertainties of the environmental process (here: the particulate matter concentrations).  

\begin{figure}
  \includegraphics[width = 0.5\textwidth]{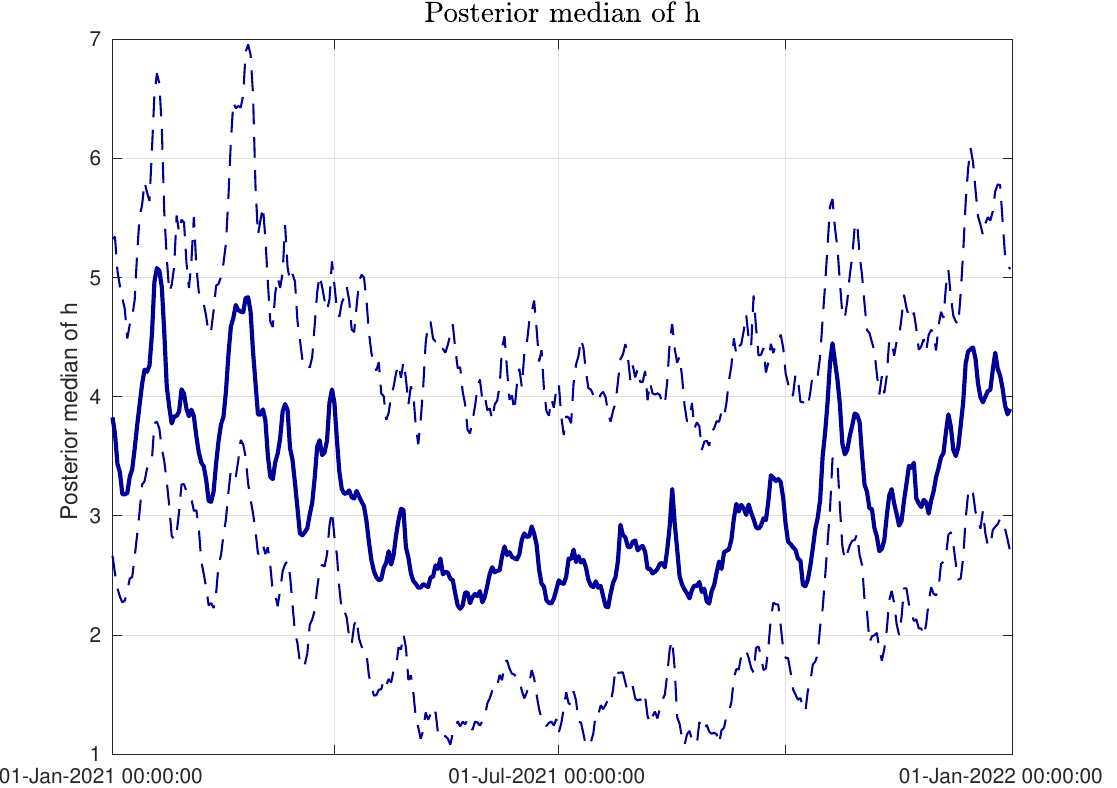}
  \includegraphics[width = 0.5\textwidth]{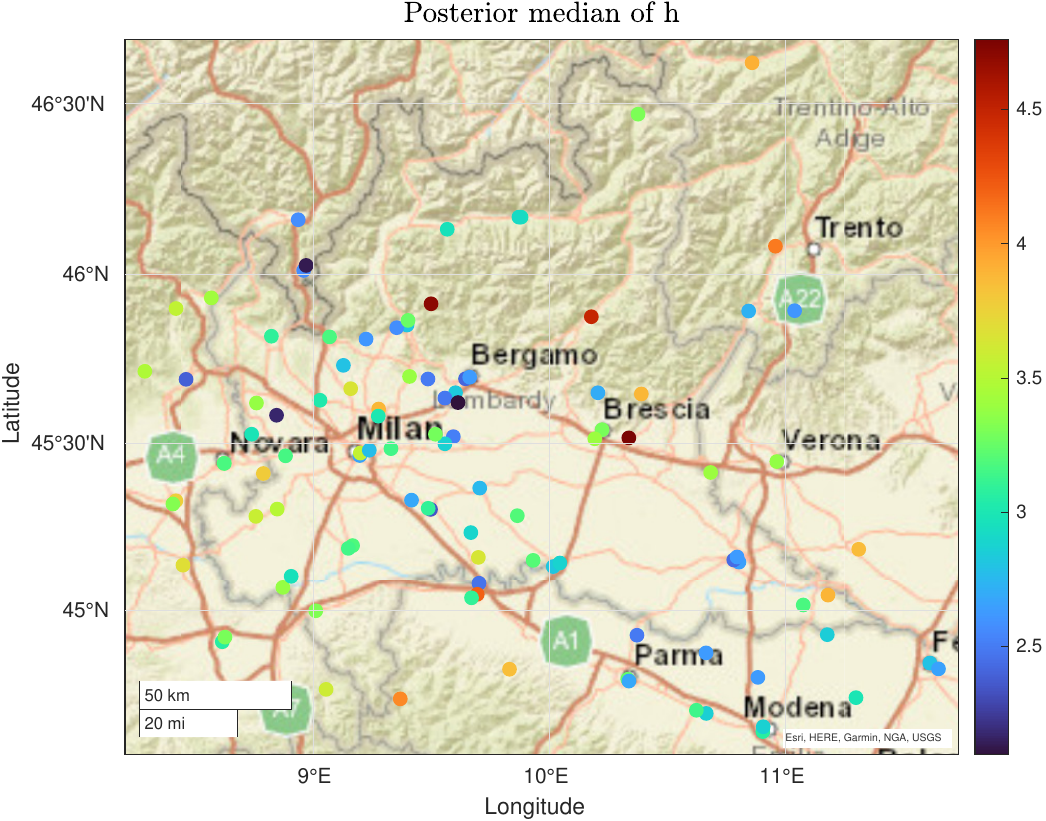}
  \caption{Median posterior draws of $\mathbf{h}_t$ (Model B) and the 5\% and 95\% quantiles across space displayed as time series (left) and median posterior draws of $\mathbf{h}_t$ across time shown on a map (right).}\label{fig:median_h}
\end{figure}

Moreover, we analyzed the conditional volatilities for each station across the time horizon. For Model B, the results are displayed in Figure \ref{fig:median_h} as a time-series plot (left) and on a map (right). More precisely, we computed the posterior medians of $h_{it}$ for all locations and time points leading to an estimate of the $n \times T$ matrix of the logarithmic conditional volatilities. Then, for visualization, we computed the median and the 5\% and 95\% quantiles of these estimates across all spatial locations (see time series plot) and across all time points (see map). From the time series plot, we see that there is a clear annual pattern -- the model uncertainties are lower in summer and highest in winter. These varying conditional volatility levels can be interpreted as environmental risks, which are the highest the winter season. In the same manner, measurement stations with a greater uncertainty/risk can be identified from the map. They are mostly located in valleys in the mountain areas.

\begin{table}
  \centering
  \caption{Estimated parameters of the dynamic stochastic volatility model, medians and 95\% posterior credible intervals.}\label{tab:svm_model}
  \begin{tabular}{ll cc cc}
      \hline\hline
    & & \multicolumn{2}{c}{Model A}  & \multicolumn{2}{c}{Model B}  \\
    & & Median & 95\% credible interval & Median & 95\% credible interval \\
      \hline
    \multicolumn{2}{l}{Intercept $\boldsymbol{\mu}$}            &         &                    &         &                     \\
    & Average constant term                                     &  3.4326 & (3.3386, 3.5274)   &  3.2774 & (3.1645, 3.3931)    \\
    \multicolumn{2}{l}{Space-time interactions}                 &         &                    &         &                     \\
    & $\rho_1$ (spatial interaction)                            &  0.5943 & (0.5547, 0.6372)   &  0.3037 & (0.2806, 0.3325)    \\
    & $\rho_2$ (temporal interaction)                           &  0.3781 & (0.3353, 0.4181)   &  0.6699 & (0.6406, 0.6939)    \\
    & $\rho_3$ (spatiotemporal interaction)                     & -0.0242 & (-0.0282, -0.0189) & -0.0236 & (-0.0279, -0.0185)  \\
    \multicolumn{2}{l}{Stochastic volatilities $\mathbf{h}_t$}  &         &                    &         &                     \\
    & Average log-volatility                                    &  3.4140 & (3.3972, 3.4313)   &  3.2299 & (3.2134, 3.2467)    \\
    \multicolumn{2}{l}{Error variance $\sigma^2$}               &         &                    &         &                     \\
    & $\sigma^2$ (variance of $\mathbf{U}_t$)                   &  0.2267 & (0.1996, 0.2510)   &  0.3312 & (0.3110, 0.3521)    \\
    \hline
  \end{tabular}
\end{table}

\section{Conclusion}\label{conc}

We have introduced a novel spatiotemporal statistical model for stochastic volatilities, which are spatially, temporally, and spatiotemporally correlated and have different levels for each location reflecting the heterogeneity of the process across space. The model includes two different error terms for the mean and the log-volatility equations. To estimate the parameters, we suggested a Bayesian MCMC approach. To this end, we applied a log-square transformation to transform our non-linear state-space model into a linear one. Further, we used a Gaussian mixture distribution to approximate the distribution of the transformed error terms in the space equation, transforming our model into a linear Gaussian state-space model. We analyzed the estimation performance for different parameter settings and spatial interactions in a simulation study, and showed that the suggested Bayesian sampler performs satisfactorily.

Moreover, we applied the dynamic spatiotemporal stochastic volatility model in a completely new empirical framework, namely, in the field of modeling environmental and climate risks. First, statistical modeling of the volatility process of climate variables have not been done extensively yet, even though there is a large scientific consensus that an increased variability of environmental processes, e.g., the temperature variability, is harmful for the environment. Second, stochastic volatility models were predominantly applied to financial data, because of the straightforward interpretation of the log conditional volatilities as the (return) risk of financial assets. We have transferred to this idea to environmental risks and showed how the volatility of $PM_{10}$ predictions are correlated across space and time. In particular, measurement stations with an increased model uncertainty could be identified in this way. In addition, we showed in our application that there are significant temporal and spatial spillovers in these environmental risks. Also, the temporally lagged spatial spillovers, i.e., the spatiotemporal correlations, appear to play a minor role.

Both from a theoretical and applied perspective, spatiotemporal stochastic volatility models and environmental risk modeling are important topics for future research. The current model does not allow for temporally varying constant terms in the conditional volatilities, which are constant across space. Regarding the latter case, other environmental processes, like temperature variations, soil droughts, or atmospheric ozone concentration and optical depths are important processes, where a deep understanding of the spatiotemporal interactions in the variabilities is essential, both for obtaining accurate prediction intervals and the planning of interventions (and their impact on the environment).





\end{document}